\documentclass[aps,pra,floatfix,superscriptaddress,nofootinbib,a4paper,final]{revtex4-2}
\usepackage{amsmath}
\usepackage{amsfonts}
\usepackage{amssymb,bm,mathbbol,txfonts}
\usepackage{siunitx}
\usepackage{graphicx}
\usepackage[usenames]{color}
\usepackage[bookmarksnumbered=true,breaklinks=true]{hyperref}
\usepackage{subfig}
\usepackage{dcolumn}
\newcommand{\eqnref}[1]{Eqn.~(\ref{#1})}		
\newcommand{\figref}[1]{Fig.~\ref{#1}}			
\newcommand{\tabref}[1]{Tab.~\ref{#1}}			
\newcommand{\secref}[1]{Section~\ref{#1}}		

\newcommand{\um}[1]{\SI{#1}{\micro\metre}}
\newcommand{\atan}{\mathop{\rm atan}}
\newcommand{\ri}{{\rm i}}						
\newcommand{\re}{{\rm e}}						
\newcommand{\g}{\gamma}

\newcommand{\vare}{\varepsilon}

\newcommand{\n}{\nu}

\newcommand{\vp}{\varphi}

\renewcommand{\Xi}{\Xi}

\def\app#1#2{%
  \mathrel{%
    \setbox0=\hbox{$#1\sim$}%
    \setbox2=\hbox{%
      \rlap{\hbox{$#1\propto$}}%
      \lower1.1\ht0\box0%
    }%
    \raise0.25\ht2\box2%
  }%
}
\def\approxprop{\mathpalette\app\relax}
\graphicspath{{./}{./figures/}}
\begin{document}
\title{\vspace{2cm}Efficient Reduction of Casimir Forces by Self-assembled Bio-molecular Thin Films\\*\emph{Preprint Version}}
\author{Ren\'e Sedmik}
\email{rene.sedmik@tuwien.ac.at}
\affiliation{\emph{on the leave from: }Department of Physics \& Astronomy and LaserLaB, VU Amsterdam (Netherlands)}
\affiliation{Atominstitut, TU Wien, Vienna (Austria)}
\author{Alexander Urech}
\affiliation{\emph{on the leave from: }Department of Physics \& Astronomy and LaserLaB, VU Amsterdam (Netherlands)}
\affiliation{Faculty of Science, University of Amsterdam (Netherlands)}
\author{Zeev Zalevsky}
\email{zeev.zalevsky@biu.ac.il}
\affiliation{Bar Ilan university, Ramat-Gan (Israel)}
\author{Itai Carmeli}
\email{itai@post.tau.ac.il}
\affiliation{Bar Ilan university, Ramat-Gan (Israel)}
\affiliation{Tel Aviv university, Tel Aviv, (Israel)}
\keywords{Casimir effect; molecular thin films; photosystem I; surface plasmons,spectroscopy}
\date{\today}
%
\begin{abstract}
Casimir forces, related to London-van der Waals forces, arise if the spectrum of electromagnetic fluctuations is restricted by boundaries. There is great interest both from fundamental science and technical applications to control these forces on the nano scale. Scientifically, the Casimir effect being the only known quantum vacuum effect manifesting between macroscopic objects, allows to investigate the poorly known physics of the vacuum. One major goal interesting from both the fundamental and the application point of view, is to understand how Casimir forces can be modified. In this work, we experimentally investigate the influence of self-assembled molecular bio and organic thin films on the Casimir force between a plate and a sphere. We find that the molecular thin films, despite being a mere few nanometers thick, reduce the Casimir force by up to 14\%. To identify the molecular characteristics leading to this reduction, five different bio-molecular films with varying chemical and physical properties were investigated. Spectroscopic data reveal a broad absorption band whose presence can be attributed to the mixing of electronic states of the underlying gold layer and those of the molecular film due to charge rearrangement in the process of self-assembly. Using Lifshitz theory we calculate that the observed change in the Casimir force is consistent with the appearance of the new absorption band and the change in the effective dielectric properties due to the formation of molecular layers. The desired Casimir force reduction can be tuned by stacking several monolayers, using a simple self-assembly technique in a solution. The small molecules -- each a few nanometers long -- can penetrate small cavities and holes, and cover any surface with high efficiency. This process seems compatible with current methods in the production of micro-electromechanical systems (MEMS), which cannot be miniaturized beyond a certain size due to `stiction' caused by the Casimir effect. Our approach could therefore readily enable further miniaturization of these devices. \nopagebreak 
\end{abstract}
\maketitle
%
\section{Introduction}
\label{sec:Intro}
The Casimir effect, predicted 1948 by Hendrick Casimir~\cite{Casimir:1948} and quantitatively confirmed 1996~\cite{Lamoreaux:1996wh} is the only pure quantum effect sourcing a force between macroscopic bodies. Even after 75 years of research, fundamental questions remain~\cite{Mostepanenko:2021cbf}. The Casimir effect, being a direct consequence of the confinement of vacuum fluctuations at (sub-)micrometer surface separations, provides a direct experimental handle to probe the quantum vacuum, which itself may hold the key to understanding dark energy -- one of the most challenging issues of contemporary physics.\\
In addition to fundamental scientific interest, manipulating Casimir forces holds great promise to enable future miniaturization of micro- and nano-electromechanical systems (MEMS and NEMS, respectively). 
In the semiconductor industry a Moore-type law existed for the feature size of transistor gates for half a century~\cite{Yeric:2015}. Given the physical limitations of fabrication processes, the industry embarks on a new trend called `More than Moore'~\cite{Waldrop:2016} -- the integration of more functionality on the same chip -- in which MEMS play an increasingly important role~\cite{Fang:2013}. The required transition from MEMS to NEMS, however, is naturally limited by quantum vacuum forces~\cite{Barcenas:2005} in some applications. The Casimir effect influences the behavior of MEMS/NEMS~\cite{Rodriguez:2015,Rodriguez:2011} and leads to `stiction' -- the unintended adhesion of surfaces by vacuum forces. In present devices, having typical surface separations of a few \um{}, the strongest impact of Casimir forces is seen near the so-called pull-in~\cite{Zhang:2014,Ardito:2012,Jia:2011} instability and via non-linear effects~\cite{Chan:2001zza,Jia:2010,Broer:2013bxa,Tajik:2018}. However, also direct applications such as Casimir-actuated devices~\cite{Chan:2001zzb,Emig:2007ea}, optical choppers~\cite{Klimchitskaya:2018}, and Casimir-driven rotary machines~\cite{Sedmik:2007,Miri:2008,Etesami:2015,Miri:2016} have been investigated. Actuatoric applications would require a way to actively control the Casimir force~\cite{Chen:2007,Sedmik:2007}. As such schemes are difficult to realize with current industrial processes and actuation via the Casimir effect would hardly yield advantages over electrostatic actuation, the baseline for research has been to reduce the Casimir force in MEMS. This has been achieved with different materials~\cite{Torricelli:2010a,Torricelli:2011,Chang:2011,Banishev:2012zza} acting as an `anti-Casimir coating' and by modifying the surface structure~\cite{Yannopapas:2010,Intravaia:2013}. Even repulsive Casimir forces have been demonstrated using liquids as intervening medium~\cite{Lee:2001,Munday:2009fgb}, and for non-trivial geometries~\cite{Tang:2017,Wang:2020ois}. In practice, however, most of the proposed methods to reduce Casimir interactions are not generally applicable, as stiction often occurs between narrow-spaced large-scale vertical surfaces created by deep trench etching~\cite{Ardito:2012}. Even if coatings could be deposited on such surfaces, they would need to be several hundred nanometers thick due to the skin depth effect~\cite{Lisanti:2005} -- thereby most probably clogging features and reducing mechanical finesse. Eventually, neither the implementation of finely structured geometries nor the deposition of materials are easily feasible in deep trenches. In the present article, we put forward a new approach that avoids these problems from the onset.\\
Self-assembled monolayers (SAMs) are thin films a mere few nanometers thick, formed by molecules equipped with a binding group on one of their ends that allows them to interact favorably with a solid surface. Well-organized and oriented monolayers \cite{Ulman:2013} can be formed in the process of self-assembly. On metallic substrates coated by a SAM of bio/organic molecules, the mixing of molecular and surface electronic states leads to the formation of a new hybrid state of matter that often presents novel electronic/magnetic properties, giving one the opportunity to investigate interesting (sometimes unexpected) physical effects~\cite{Carmeli:2003,Hernando:2006a,Hernando:2006,Vilan:2000a,Dintinger:2005,Canaguier-Durand:2013,Munday:2009fgb,Carmeli:2012a,Esat:2017,Chen:2010,Garcia-Vidal:2021b,Orgiu:2015a,Carmeli:2003a,Neuman:2006,Carmeli:2023}. Among others, it has been demonstrated that bio/organic coatings can dramatically extend molecular coherent states and induce Rabi splitting. These phenomena ultimately give rise to significant changes in the reflection, transmission, and absorption spectra of the surface~\cite{Carmeli:2012a,Dintinger:2005,Canaguier-Durand:2013,Chen:2010,Garcia-Vidal:2021b}. 
In some cases, the interaction of molecular-plasmon hybrid states with electromagnetic vacuum fields can change material properties such as charge and energy transfer, magnetism, or chemical reactivity~\cite{Garcia-Vidal:2021b,Orgiu:2015a}. It was also shown that the adsorption of molecular layers can result in new electronic properties of the surface, which emerge due to electron transfer between the adsorbed monolayer and the surface, causing substantial charge rearrangement at the surface level. This charge rearrangement was observed as a new absorption band of a gold/monolayer surface~\cite{Carmeli:2003a} and was attributed to delocalized electrons, suggesting the appearance of plasmon or exciton-like behavior of the surface electrons. The band correlated with the level of organization of the molecular film and its molecular dipole moment~\cite{Neuman:2006}. For instance, when a monolayer of molecules with a large dipole moment self-assembles on top of a metal, the electrons on the surface of the metal undergo a substantial rearrangement, giving rise to a marked change in electron density~\cite{Carmeli:2003,Hernando:2006a,Hernando:2006, Orgiu:2015a,Carmeli:2003a,Neuman:2006}, and thus, in the effective dielectric function of the surface in a wide frequency band. As Casimir interactions strongly depend on the dielectric properties of the confining boundaries, the coating of surfaces by specific SAM could therefore provide a new way of reducing Casimir forces.

In the present work we investigate the Casimir interaction between a gold-coated sphere and flat substrates carrying monolayers of Photosystem I (PSI,  single and double layer), polypeptide Poly Proline in PPI conformation (PPI), Poly Proline in PPII conformation (PPII) and the alkylthiol 1-octadecanethiol (C18). All films are self-assembled atop a thick layer of gold. Using a custom-built atomic force microscope, we demonstrate a reduction of up to 5\% in the force per monolayer of only a few nm thickness and maximum reduction of 14\% for a double layer. 
The self-assembly property of the investigated molecules could be used, to form a well-specified number of homogeneous layers on internal surfaces of deep trenches im MEMS. We believe that such coatings could readily be implemented in industrial production processes and therefore represent a viable practical way to reduce Casimir forces by a significant amount wherever needed.\\
In the following, we first discuss the basic properties of the investigated molecules in \secref{sec:films}.  
Our experiment and its results are discussed in \secref{sec:casimir}, followed by a detailed comparison between theory and experiment. We summarize our findings in~\secref{sec:conclusion}.
%
\section{Self-assembly of single and multi-layer thin films}
\label{sec:films}
\begin{figure}[!h]
    \centering
    \includegraphics[width=110mm]{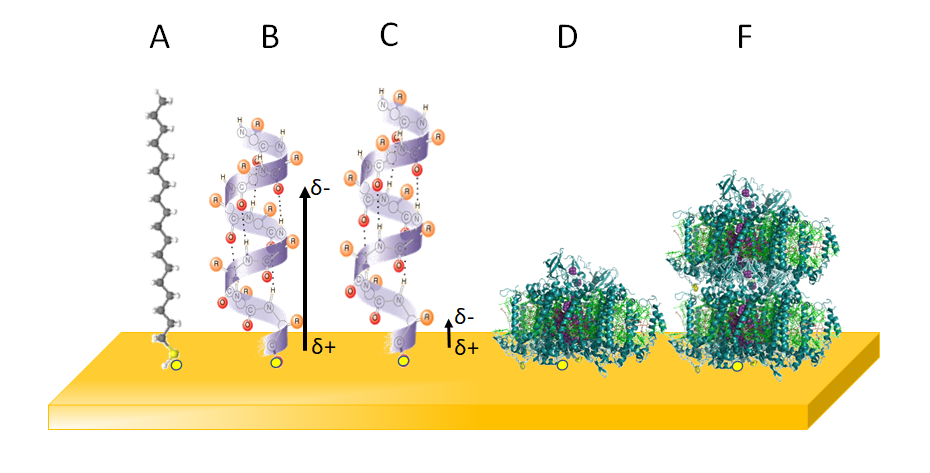}
    \caption{Schematic illustration (not to scale) of the four types of molecules and five films used in this study.  1-Octadecanethiol (C18) (A) Polyproline in Cis (PPI) (B) and Trans (PPII) (C) confirmations. Single (D) and double (F) PS I layers. The arrows illustrate qualitatively the difference in dipole moment strength between PPI and PPII. All molecules bind covalently to the surface by a thiol group (space fill model - yellow). \label{fig:molecules}}
\end{figure}

Five different types of Organized Organic Thin Films (OOTFs) were introduced in this study. We chose the films so that they allow to test a large dynamical range of basic molecular properties, such as molecular dipole moment, conduction coefficient, and film thickness. \figref{fig:molecules}  illustrates schematically the different types of molecular films utilized in the study.  

All used monolayers bind covalently to the metal surface through a thiol group and form oriented and organized SAMs. The first type of layer (C18) is composed of alkylthiols 1-Octadecanethiol (Sigma-Aldrich) containing 18 carbons of linear chain with a thiol group at one end. This molecule has a small dipole moment of $\sim 2$ Debye~\cite{Ramirez:2007} and a low real dielectric constant at low frequencies: $\varepsilon(0)=2.1$ \cite{Szwajca:2015}. C18 is highly hydrophobic and forms a very thin ($\sim2\,$nm) and well organized film on Au surfaces \cite{Ishida:1997,Ulman:1996}.

The second film is composed of a helical, 36 amino acids long, polypeptide Poly Proline (PP) with Cysteine (Cys) containing a thiol group at the C-Terminal CysP36 (Genemed Synthesis, Inc). The peptide based on PP was chosen due to PP’s special and relatively rigid structural properties and its ability to adopt two conformations PPI and PPII depending on the solvent of choice. PPI has a CIS conformation with the amide bonds oriented nearly parallel to the helix axis~\cite{Kuemin:2009}. The helix is only stable in low polarity solvents such as n-propanol. The form observed in aqueous solution and in more polar alcohols is PPII, an all-trans conformation. Within this helix, all amide bonds of the peptide backbone are nearly perpendicular to the helix axis. In the monolayer, the electric dipole moment of PPI is perpendicular to the surface~\cite{Adzhubei:2013,Han:2011,Holzwarth:1969,Kuemin:2009} 
while that of PPII is parallel to the surface. The ability of the peptide to adopt two conformations allows us to study the effect of the orientation of the molecular film dipole moment on Casimir forces. The dipole moment along the helix axis of PPI helices is predicted to be about 5 times larger as compared to that of PPII Helixes \cite{Kuemin:2009}. For example, PP composed of fifteen units in PPI conformation will have a dipole of 70 Debye compared to 15 Debye in the PPII conformation. The peptide length ranges from $\sim6$--$9\,$nm, depending on the confirmation of the PP. Its dielectric constant $\varepsilon(0)=2.3$ is a bit higher than that of 1-Octadecanethiol. AFM images (see below) confirm the formation of dense monolayers for both PPI and PPII with average thicknesses of $4\pm0.4\,$nm and $5.2\pm0.4\,$nm, respectively, as recorded by Ellipsometry.

The fourth and last type of film is composed of the photosynthetic protein Photosystem I (PS\,I). The PS\,I used in this study, derived from cyanobacterial membranes, is a helical protein membrane complex composed of 96 chlorophyll and 22 carotenoid pigment molecules. It has a cylindrical shape with a diameter of \SI{15}{\nano\metre} and a height of \SI{9}{\nano\metre}. It harvests photons with a quantum efficiency of 1 in the visible range\cite{Brettel:1997,Gerster:2012,Carmeli:2015}, and was found to be stable and photoactive when adsorbed in a dry environment. We fabricated PS\,I SAMs by formation of a direct sulfide bond between unique cysteine mutants in PS\,I and the patterned Au surface. The mutations D235C/Y634C are selected near the special chlorophyll pair P700 to allow close proximity between the reaction center and the gold substrate. Our self-assembly approach facilitates efficient electronic junctions and avoids disturbance in the function of the reaction center~\cite{Gerster:2012}. The PS\,I was chosen for its high dielectric response at DC $\varepsilon(0)=7$~\cite{Carmeli:2015} and its ability to interact efficiently with electromagnetic radiation through its extended antenna system. Another advantage of PS\,I is the ability to form multilayers by crosslinking~\cite{Carmeli:2015}. In order to investigate the effect of multiple layers on the Casimir effect, we fabricated samples with two PS\,I layers by covalently binding PS\,I to existing layers using an m-maleimidobenzoyl-N-hydroxysulfosuccinimide ester (Sulf-MBS) cross linker. The approach allows to investigate how an increase in the film thickness affects Casimir forces.     

All monolayers were chemisorbed onto a \SI{200}{\nano\metre} gold layer above a \SI{2}{\nano\metre} Ti adhesion layer on mica surfaces (SPI supplies, grade V1). For the fabrication of the C18 and two PP films we first cleaned the gold surface in ozone for 50 minutes and then one hour in pure Ethanol (Merck). The samples were then washed with the pure EtOH and dried in N$_2$. Self-assembly was carried out in \SI{1}{\milli\mol} Chloroform (Merck), n-propanol, and deionized water solutions for the C18, PPI and PPII molecules, respectively overnight. All samples were then washed in the pure adsorption solvents and dried in N$_2$. Adsorption of PS I was performed as described previously \cite{Gerster:2012,Carmeli:2015} from a \SI{20}{\milli\mol} PBS buffer solution. In short, the first step was formation of self-assembled mono layers of PS I by a direct sulfide bond between unique cysteine mutants in PS I and the Au surface. Adsorption of the PS I monolayer was achieved by first cleaning the Au surface in an Ar and O$_2$ (50\%:50\%) plasma. The surface was then washed in EtOH for 2 min., dried and placed in a concentrated solution of PS\,I for 2 h, followed by an additional washing step using buffer and deionized water. The multilayer was fabricated by cross linking of successive PS\,I layers with  Sulf-MBS (Pierce). This technique ensures that the PS\,I multilayer is covalently bound to the surface in an oriented fashion with a good electronic coupling to the underlying gold \cite{Heifler:2018}. 
All samples were kept at room temperature in N$_2$ until insertion into the apparatus for force gradient measurements or ellipsometry, where they were exposed to ambient air.
%
\section{Reduction of Casimir forces}
\label{sec:casimir}
In order to assess the reduction of the Casimir effect by SAMs, we measured the force gradient between a gold-coated sphere and either a flat gold substrate or a gold substrate carrying a self-assembled bio-molecular monolayer as described above. In the following, we describe the setup and the measurements performed with it.
%
\subsection{Experimental setup}
\label{sec:casimir:exp}
\begin{figure}[!ht]
 \centering
 \includegraphics{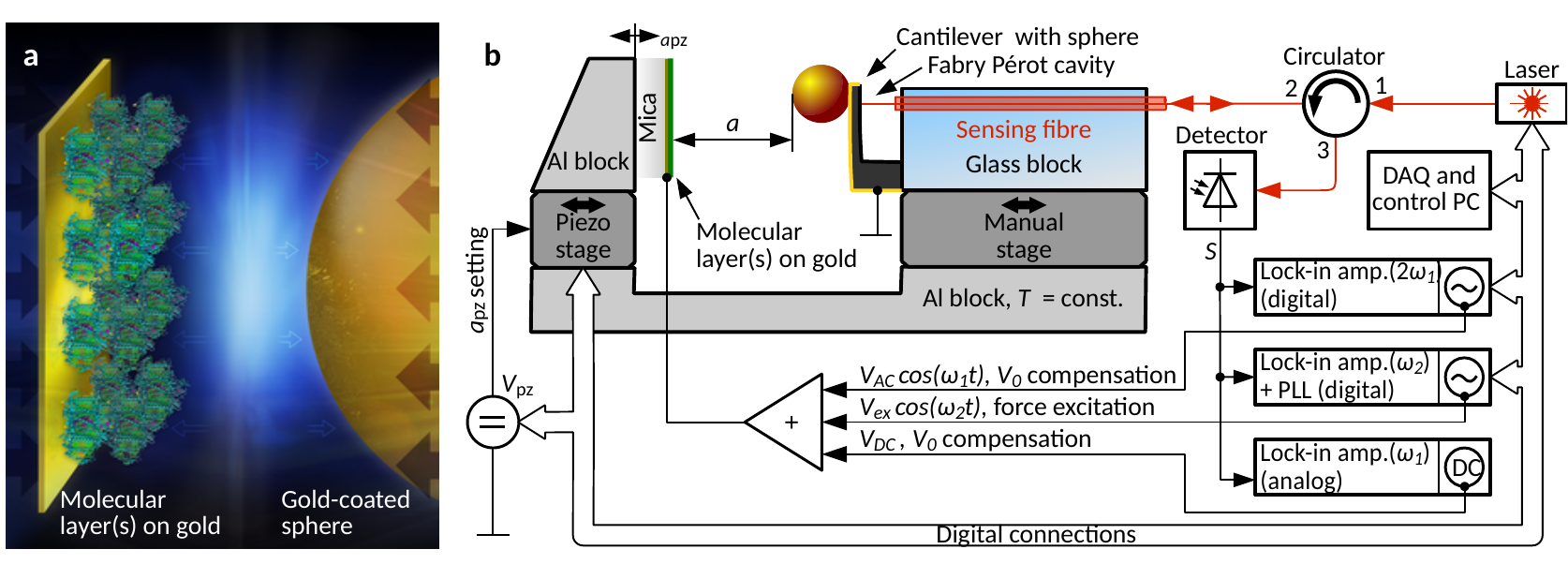}
 \caption{Experimental setup. \textbf{a)} Artist's impression of the interaction region showing the monolayers on top of the gold substrate opposing the gold sphere. \textbf{b)} Schematic of the setup defining the different potentials and feedback circuits (see main text). \label{fig:setup}}
\end{figure}
We used a modified version of a custom-made atomic force microscope (AFM) described in the literature~\cite{Zuurbier:2011sy}, and shown in \figref{fig:setup} (details are given in the methods section). On the right hand side, a \um{158} diameter polystyrene sphere is mounted to the flexible end of a silicon nitride cantilever and sputter coated with $>100\,$nm gold atop a chromium adhesion layer, creating a mass-spring system with effective mass $m$ and free resonance frequency $\omega_0$. We use the reflective back side of the cantilever to form a Fabry~P\'erot cavity with the cleaved end of an optical fiber placed roughly \um{500} from it. This configuration allows us to detect the movement of the cantilever and its response to forces acting between the sphere and the flat plate (on the left side in the figure). The latter carries the actual subject of this investigation: various molecular monolayers atop an optically thick gold film on a mica substrate. The entire setup is placed inside an aluminum enclosure on top of an active vibration-isolating table inside an anechoic box.\\
We perform measurements of the force gradient $\partial F(a)/\partial a\equiv \partial_a F$ between the sphere and the plate by detecting the shift in the cantilever's resonance frequency. Measurements are conducted in sweeps during which the separation $a$ is reduced from $500\, $nm to $\sim70\,$nm in 34 logarithmically spaced steps. At each position, before taking data, several independent feedback circuits are given at least 6 times their respective time constants in order to swing in and compensate the following systematic effects: First, the inevitable surface potential~\cite{Speake:2003zz} $V_0$ is actively compensated using a modulation technique~\cite{deMan:2009,Sedmik:2013} that aims to minimize the global electrostatic force between the two surfaces by adapting a DC bias voltage. Second, the signal amplitudes due to $V_{AC}$ ($V_0$ compensation) and $V_{ex}$ (resonance excitation) are kept constant by adapting the respective voltages. Thereby, we ensure an equal signal-to-noise ratio for the $V_0$-feedback circuit and the frequency measurement over the whole range of distances. Third, the temperature of the rigid aluminum base and enclosure (not shown) is controlled with a precision of less than $10\,$mK at a temperature of roughly $296\,$K. Small fluctuations in the ambient air's temperature and relative humidity, as well as long term drift of the glue used to mount the sphere on the cantilever still result in small changes of the working point of the cantilever and piezo drift. In order to compensate for these effects, a fourth feedback circuit adapts the interferometer's laser wavelength such that the signal stays at the same (quadrature) point of the interferometric fringe, guaranteeing a constant sensitivity of the optical detection system.\\
At each step, the calibrated piezo position $a_{pz}$ is recorded. The separation $a=a_0-a_{pz}$ for all approached positions is determined and corrected \emph{a posteriori} after each sweep using the recorded electrostatic signal~\cite{deMan:2009}. Finally, the gradient 
\begin{align}
 \frac{\partial F(a)}{\partial a}=m\left[\omega_0^2-(\omega_0+\Delta\omega)^2\right]\,,
 \label{eq:force_gradient}
\end{align}
of the force $F(a)$ between the sphere and the plate is detected via a phase-locked loop (PLL) sensing the frequency shift $\Delta \omega=\omega_0(\partial_a F)-\omega_0(0)$, similar as described in the literature~\cite{Kleiman:1985,Albrecht:1991,Decca:2003zz,Sedmik:2018kqt}. In contrast to these investigations, however, we do not place our setup in vacuum but keep it in air. For this reason, proper choice of the excitation mechanism (electrostatic, via the second harmonic of an AC voltage signal $V_{ex}$ at frequency $(\omega_0-\Delta \omega)/2$) and the calibration of the phase setpoint $\phi_{set}$ for the phase-locked loop (PLL) are crucial, as is detailed in the methods section. For each sample, we first re-calibrate $\phi_{set}$ by recording the cantilever response as a function of frequency at $a=a_{cal}=\SI{2.5}{\micro\meter}$ and subsequent least-squares fitting of the measured phase to the cantilever transfer function. Then, we perform roughly 35 distance sweeps, each preceded by a calibration of $\omega_0$ at $a_{cal}$. Finally, the effective mass $m$ of the cantilever is determined by least squares fitting of $\partial_a F_{ES}$ as a function of a DC potential $V_{DC}$ applied to the plate at known separation.\\ 
%
\subsection{Experimental Results}
\label{sec:casimir:res}
\begin{figure*}[!ht]
 \centering
 \includegraphics[width=\textwidth]{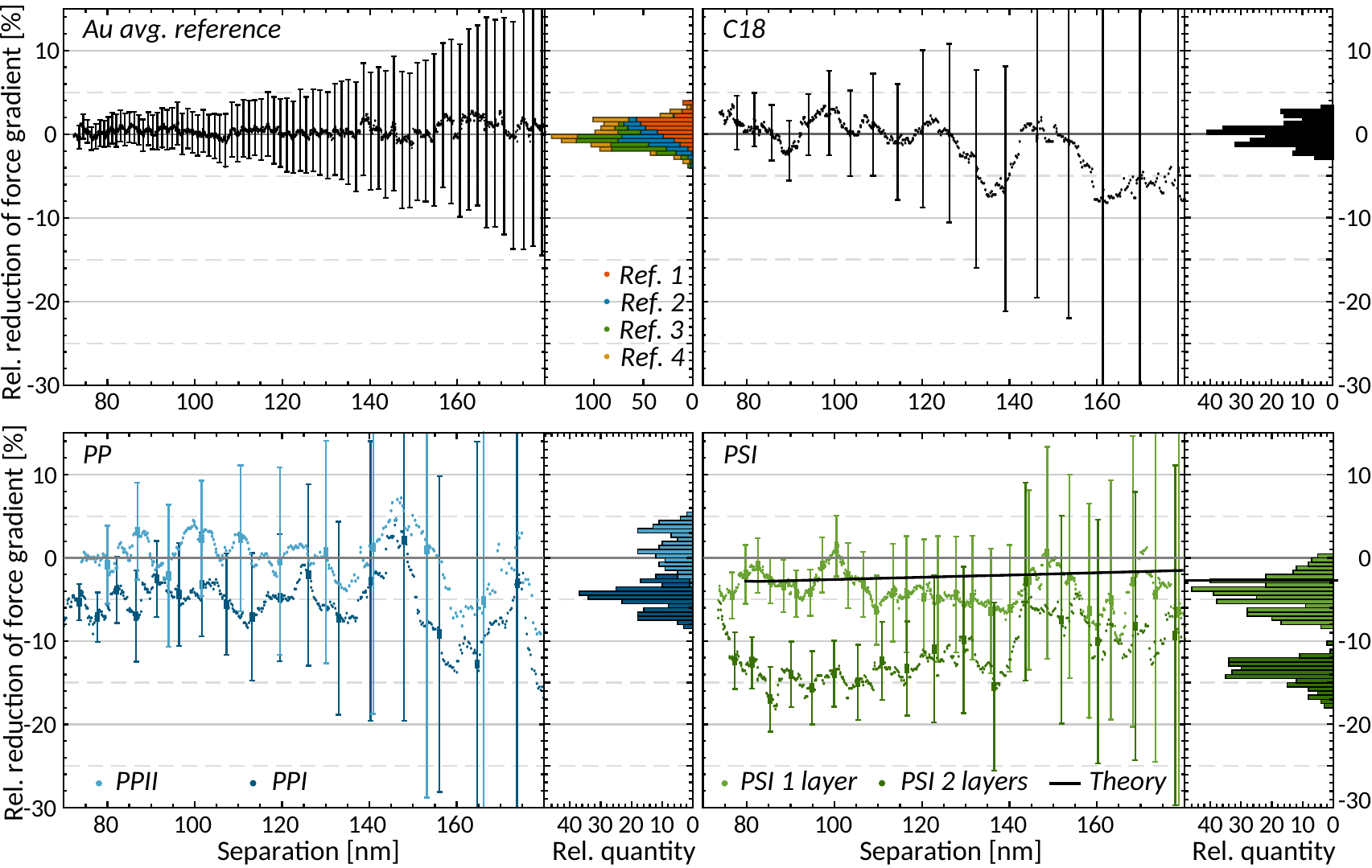}
 \caption{Reduction of the measured Casimir force gradient between a gold coated sphere and flat gold substrates with various molecular layers on top as a function of surface separation (left graphs), relative to the reference (gold vs. gold). For clarity, only every 30$^\text{th}$ error bar is shown. Note that data have been averaged (see main text). Histograms for the range 80--120$\,$nm are attached right of the respective data. For the references, an overlay of the histograms of all four reference measurements is shown. For PS\,I, the theoretical expectation including the measured dielectric properties, roughness, and electrostatic patches is included as well as the average reduction in the range 80--120$\,$nm as black line in the histogram.\label{fig:results}}
\end{figure*}

We measured the force gradient between a gold-coated sphere and flat substrates coated with monolayers of C18, PS\,I, PPI, PPII, as well as a substrate with two layers of PS\,I. 
In order to highlight the effect of the molecular layers on the Casimir force, we eliminate small residual systematic effects~\footnote{One systematic effect is the increasing non-linearity of the cantilever's response at the shortest separation, resulting in an up-bending of the force gradient curve in this regime. Another effect is ringing of the PLL circuit at specific separations due to imperfect adaptation of the feedback to the highly non-linear force gradient.} by comparing in \figref{fig:results} only the relative differences between measured force gradients on substrates coated with various bio-molecular layers and the force gradient measured with the same sphere on a flat gold substrate. As zero reference, we use a weighted moving average over the weighted mean of four different reference measurements~\footnote{We have performed four different measurements (top left in the figure) between the same sphere and two different gold reference samples, before, between, and after measurements on coated substrates. These references agree within the experimental standard deviations and do not show any trend over time, which serves as a check for the surface integrity of the sphere.}. Histograms for the separation range 80--120$\,$nm are given on the right of the respective gradient plots in \figref{fig:results} for the single references (top left) as well as for all measurements on substrates. We chose this limited range, as at shorter and larger separations, the effects of roughness and noise, respectively, dominate.

Numerical values for the weighted mean reduction of each sample are collected in \tabref{tab:reduct} together with the measured thickness, roughness, and surface potential deviations of the bio-molecular films.
\begin{table}[!ht]
\caption{Weighted mean and standard deviation of the measured relative reduction of Casimir force gradients for each material in the separation regime 80--160$\,$nm. Additional information on the measured roughness and patch potential distributions is given.\label{tab:reduct}}
 \begin{tabular}{l r@{.}l r@{.}l r@{.}l r@{.}l r@{.}l}
 \hline\hline
  & \multicolumn{2}{c}{force gradient} & \multicolumn{2}{c}{film} & \multicolumn{4}{c}{surface roughness} & \multicolumn{2}{c}{avg. patch}\\
  sample& \multicolumn{2}{c}{rel. reduction\,[\%]}&\multicolumn{2}{c}{thickness\,[nm]}&\multicolumn{2}{c}{rms\,[nm]}&\multicolumn{2}{c}{peak-peak\,[nm]}&\multicolumn{2}{c}{potential\,[mV]}\\
  \hline
  C18         &  $\phantom{0}0$& $0\phantom{0}\pm1.4$& $\phantom{.0}$1&  $9\pm0.2$ & \mbox{}\hspace{2.5ex}3&2 & \mbox{}\hspace{4.5ex}18&3 & \mbox{}\hspace{5ex}4&7\\
  PPII  &  $\phantom{.0}1$& $3\phantom{0}\pm2.1$& $\phantom{.0}$5&$2\pm0.4$ & 3&1 & 68&8 & 1&8\\
  PPI     & $\phantom{.0}-4$& $8\phantom{0}\pm1.5$& $\phantom{.0}$4& $0\pm0.4$ & 3&5 & 37&4 & 1&5\\
  PSI 1 layer & $\phantom{.0}-4$& $0\phantom{0}\pm1.7$& $\phantom{.0}$9& $0\pm1.4$ & 2&2 & 34&7 & 1&6\\
  PSI 2 layers&            -13&  $8\phantom{0}\pm1.4$& \multicolumn{2}{c}{{\mbox{}\hspace{-11pt}$20\phantom{.0}\pm 2\phantom{.0}$}}& 3&2 & 107&0 & 1&6\\
  \hline
  Au reference & 0&$00\pm0.18$& \multicolumn{2}{c}{>200}& 1&9& 7&5 &2&4\\
  Au sphere & \multicolumn{2}{c}{-} & \multicolumn{2}{c}{>100} & 14&5 & 133&5 & 3&5\\
  \hline\hline
 \end{tabular}
\end{table}
In our experiment the C18 film was used as a control measure since it has a small influence on the dielectric response of the gold substrate in the optical region and a vanishing influence in the infrared~\cite{Prato:2008}, for which it was not expected to influence the Casimir force. We thus used this measurement as an additional check of our zero reference, and to demonstrate the absence of strong systematic effects due to (trapped) electrostatic charges on the coated surfaces. 
To study the influence of molecular confirmation and dipole effects on the Casimir force, two PP films were used which differ significantly in their dipole moment strength. We observe that while PPII barely shows any reduction, a single layer of PPI results in roughly 5\% reduction. These results suggest that the confirmation of the polypetide and its dipole moment can significantly affect Casimir interactions. However, the stark differences of the reduction for different molecules also highlight the need for further investigation of the complex interplay between molecular confirmation/dipole moment and the underlying gold layer.  

As discussed above, PS\,I was chosen due to its high dielectric constant and extended antenna system facilitating efficient interaction with photons. We have measured a reduction of $4$\% in the Casimir force for a single PS I monolayer. In \secref{sec:comparison} we show that this result agrees well with expectations from Lifshitz theory on the basis of the measured dielectric properties, roughness, and electrostatic effects. We note that on another sample (not presented here) coated with a monolayer of PS\,I that was prepared in a similar way to the samples used in the present investigation, we obtained $6.5\pm2.1$\% reduction that, however, seemed to lessen over time as the sample was exposed to air. A third similar sample showed perfect agreement with the 4\% presented here. Changes in the spectral responses of biomolecular monolayers over time in air seem to be caused by hydrocarbon adsorption~\cite{Hattori:2021}. Adding another monolayer of PS\,I increases the reduction to $\sim$14\% in the range 80--120$\,$nm. It is notable that the observed reductions in the Casimir force gradient by biomolecular monolayers are much stronger than those achievable with oxides of the same thickness. For example, a \SI{4}{\nano\metre} layer of indium tin oxide (ITO) only reduces the interaction by 1.86\%, less than half of the effect of a PPI monolayer. A thick layer of ITO was demonstrated to half the Casimir force~\cite{deMan:2009zz}.
%
\subsection{Detailed comparison with theory}
\label{sec:comparison}
\begin{figure}[!ht]
    \centering
    \includegraphics[width=\textwidth]{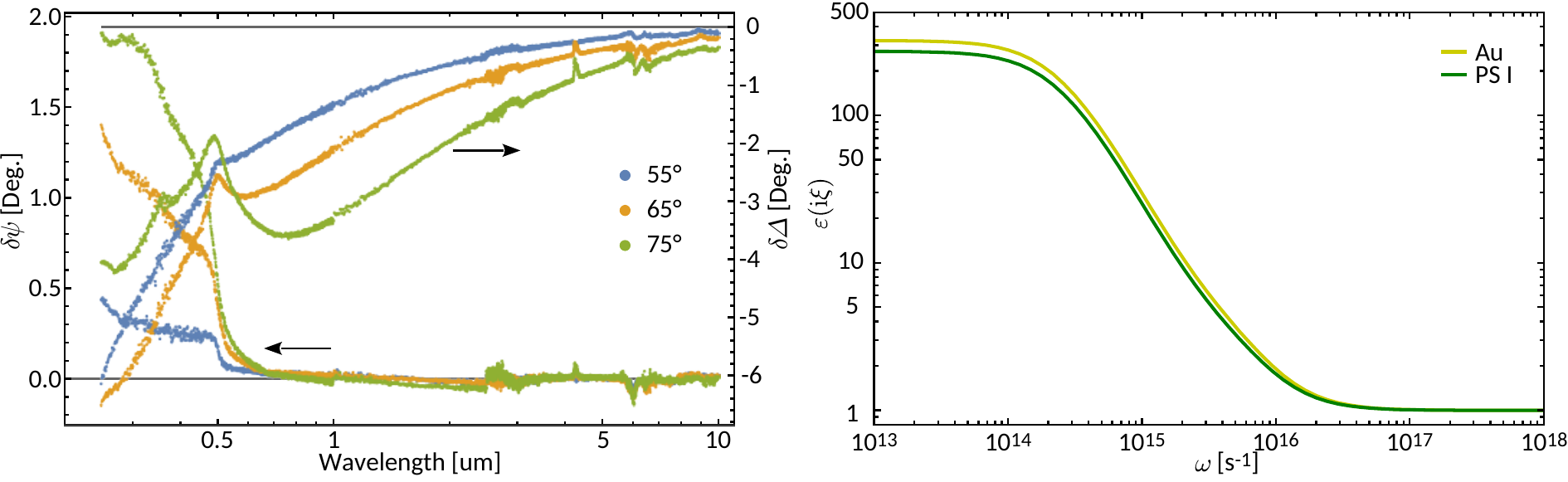}
    \caption{Spectroscopic data for the Au reference and PS\,I. Left: Differential spectra $\delta\psi(\lambda)$ and $\delta\Delta(\lambda)$ between the gold reference and the PS\,I sample obtained from ellipsometric measurements at three different angles of incidence. Right: Kramers-Kronig-transformed dielectric spectra depending on frequency $\omega$.\label{fig:spectra}}
    \label{fig:my_label}
\end{figure}

As a test case, we performed on one of the films -- the single PS\,I layer -- a precise comparison between the measured force gradient reduction and the theoretical expectation, based on the spectral dielectric response, surface roughness and electrostatic patches. Before describing the procedure in detail, we remark that the roughness is higher on all tested molecular films than on pure gold. RMS surface potentials are found to be very similar to gold for all films. As roughness and electrostatic forces both \emph{increase} the force (gradient), while we observe a \emph{reduction} thereof, these effects thus cannot invalidate our results, qualitatively, under any circumstances.\\
The complex dielectric functions of a gold substrate and a PS\,I monolayer on gold were determined in the wavelength range \SI{250}{\nano\metre}--\SI{8}{\micro\metre} using ellipsometry (see methods section). For shorter wavelengths, we appended both datasets by literature values for gold~\cite{Palik:1998a}. We then synchronously fitted the real and imaginary parts of the dielectric spectra depending on frequency $\omega$ to the Drude-type model
\begin{align}
    \vare(\omega)=1-\frac{\omega_p^2}{\omega(\omega+\ri/\tau_D)}+\sum\limits_{i=1}^{N}\frac{\xi_i}{\omega_j^2 - \ri \omega \gamma_i - \omega^2}\,,\label{eq:Drude}
\end{align}
with free parameters $\omega_p$, $\tau_D$, and several oscillator terms with strength $\xi_i$, frequency $\omega_i$, and damping $\gamma_i$, respectively. Both data and fitted models are shown in \figref{fig:spectra}. Details on the fit are given in the methods section. The resulting model was used to extrapolate data towards both lower and higher frequencies as required for the Kramers-Kronig transform~\cite{Landau:2013}, yielding $\vare(\ri \xi)$ as functions of complex frequencies~\footnote{Note that for the present comparison, the open question about the inclusion of dissipation or non-locality in the dielectric model~\cite{Mostepanenko:2021cbf} is of no concern.}. The latter were inserted into the well-known Lifshitz expression
\begin{align}
\frac{\partial F_{C}(a,T)}{\partial a}=R\frac{k_B T}{\pi c^3}{\sum\limits_{n=0}^\infty}{'}\! \int\limits_1^\infty \! p\xi_n^3 \,{\rm d}p &\Bigg\{ \Bigg[\frac{\re^{2a p\xi_n/c}}{\left[r^{(1,2)}_{TM}(p,\ri\xi_n) r^{(2,3)}_{TM}(p,\ri\xi_n) \right]}-1\Bigg]^{-1}+\Bigg[\frac{\re^{2a p\xi_n/c}}{r^{(1,2)}_{TE}(p,\ri\xi_n) r^{(2,3)}_{TE}(p,\ri\xi_n)}-1\Bigg]^{-1}\Bigg\}\>,\label{eq:lifshitzT}
\end{align}
to compute the Casimir force gradient $\partial F_{C}/\partial a$ between the plate and sphere in the proximity force approximation considering the sphere radius $R$, and ambient temperature $T=296\,$K.
The prime on the sum indicates that the term with $n=0$ receives an additional factor $1/2$ (for details see e.g.~\cite{Bordag:2014}).
We can write the (Fresnel) reflection coefficients~\footnote{The common notation for the real and imaginary parts of the complex dielectric permeabilities is $\vare(\omega)=\vare'(\omega)+\ri \vare''(\omega)$, and similar for the complex magnetic permeabilities $\mu(\omega)$.} as 
\begin{align}
\label{eq:Fresnel}
 r^{(m,m')}_{TM}(p,\ri\xi)=\frac{\vare^{(m)}(\ri\xi)\kappa^{(m')}(p,\ri\xi)-\vare^{(m')}(\ri\xi)\kappa^{(m)}(p,\ri\xi)}{\vare^{(m)}(\ri\xi)\kappa^{(m')}(p,\ri\xi)+\vare^{(m')}(\ri\xi)\kappa^{(m)}(p,\ri\xi)},\ \text{and } r^{(m,m')}_{TE}(p,\ri\xi)=\frac{\mu^{(m)}(\ri\xi)\kappa^{(m')}(p,\ri\xi)-\mu^{(m')}(\ri\xi)\kappa^{(m)}(p,\ri\xi)}{\mu^{(m)}(\ri\xi)\kappa^{(m')}(p,\ri\xi)+\mu^{(m')}(\ri\xi)\kappa^{(m)}(p,\ri\xi)},
\end{align}
where we use indices $m=1,\,3$, for the plate and sphere, respectively, and $m=2$ for the air gap with $\varepsilon^{(2)}(\ri\xi)=1$. Furthermore, we have the functions
\begin{align}
 \kappa^{(m)}(p,\ri\xi)=\sqrt{p^2-1+\vare^{(m)}(\ri\xi)}\>.
\end{align}
Note that the relative magnetic permeabilities $\mu^{(m)}(\ri\xi)=1$ are (as usual for non-magnetic materials) ignored by setting their value equal to the one for vacuum.\\
The spectra in \figref{fig:spectra} show that the PS\,I layer strongly changes the dielectric response of the substrate at all wavelengths below \SI{10}{\micro\metre}. While this broadband effect is the reason, why the computed $\varepsilon(\ri\xi)$ spectra (and hence the Casimir force) are influenced, we note that there are isolated resonances near \SI{6.0}{\micro\metre}, \SI{6.5}{\micro\metre}, and \SI{7.9}{\micro\metre}, where the changes in the dielectric response are even more pronounced.
\begin{figure}[!ht]
\centering
\includegraphics[width=\textwidth]{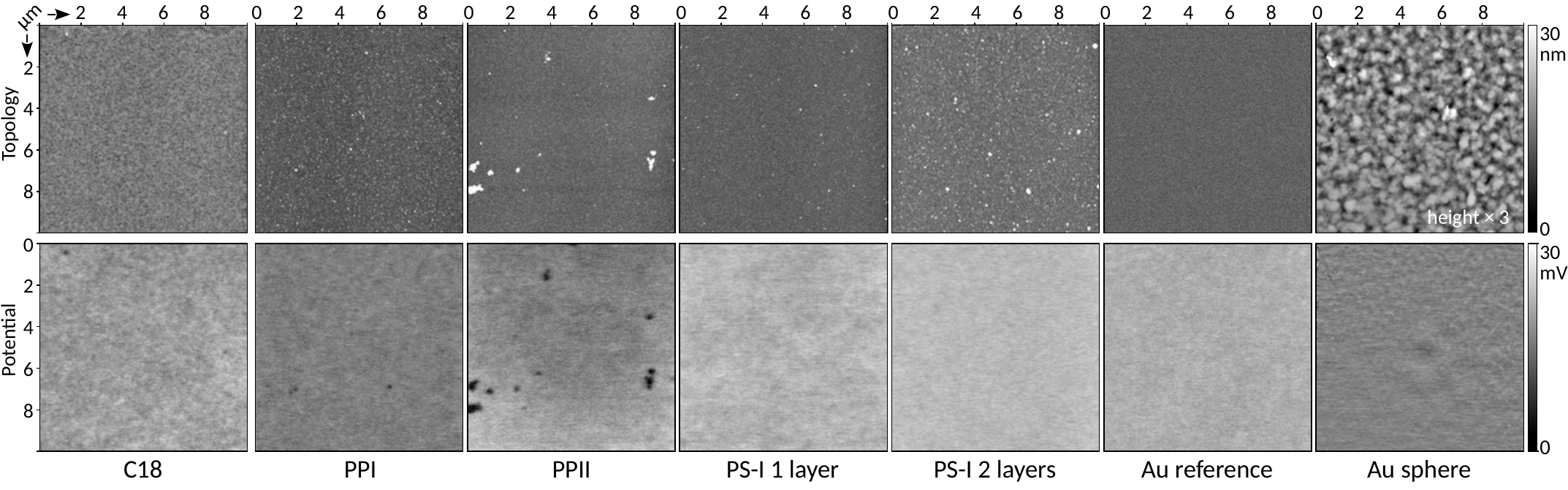}
\caption{Surface topology and potentials for all investigated substrates and the sphere, obtained by AFM and KPFM, respectively on the same spot for each surface.\label{fig:topopot}}
\end{figure}
We have performed AFM and Kelvin probe (AM-KPFM) measurements after the force gradient measurements on all samples and on the sphere, as shown in \figref{fig:topopot}. The topography of all flat surfaces shows homogeneously formed films with RMS roughness amplitudes below \SI{4}{\nano\metre}. However, the PP and PS I samples have isolated peaks extending several tens of nm. Peaks also appear at the respective positions in the measured surface potentials. We may only speculate about the nature of these anomalies but they may be caused by interlinked molecules clumping at the surface~\cite{Suresh:2022,Oliveira:2001,Ulman:1996}
, for which they might be avoidable by an optimized adsorption procedure. In the experiment, the peaks were evaded by selecting a spot at which the separation $a$ could be reduced to below \SI{50}{\nano\metre} without snap to contact. For this reason, we exclude peaks larger than \SI{30}{\nano\metre} in the analysis. The sphere shows a significantly higher roughness profile (note that the sphere topography in \figref{fig:topopot} is shown with a factor $1/3$). The peak-peak amplitude of \SI{134}{\nano\metre} is mainly caused by holes between grains of gold and isolated smaller peaks away from the tip of the sphere. Statistical values on topography and potentials are listed in \tabref{tab:reduct}. In order to compare our measurements to Lifshitz theory, we first compute $\partial_a F_C$ using the actually measured dielectric responses as described above. We then correct for roughness effects by computing the change in the Casimir force gradient for random relative positions of the sphere and plate, based on the actual AFM profiles in the PFA, and subsequent averaging, as is described in the methods section. 
Finally, patch effects are computed using a similar Monte-Carlo scheme~\cite{Behunin:2012zr} on the basis of our KPFM measurements, and all corrections are added to $\partial_a F_C$ to yield an estimation of the measured force gradient $\partial_a F_\text{tot}$. Eventually, these estimates give rise to a change $\Delta_\text{PS\,I}[\partial_a F_\text{tot}](a)=\partial_a F_\text{PS\,I}/\partial_a F_\text{Au}-1$ shown as solid line in the bottom right panel of \figref{fig:results}. In the range 80--120$\,$nm, the theoretical estimate of
$\Delta_\text{PS\,I}[\partial_a F_\text{tot}](a)=-2.7${\raisebox{0.5ex}{\tiny$\substack{+1.2 \\ -1.7}$}\% compares well with the measured reduction $-4.0\pm1.7\%$. We note that without roughness and electrostatic corrections, we obtain a theoretical reduction of $\Delta_\text{PS\,I}[\partial_a F_\text{tot}](a)=-3.24$\%

\subsection{How the force is influenced by the films}
\label{sec:discussion}
It is well known that interactions between plasmons of metal surfaces and organic molecules can give rise to a number of phenomena. Among them are refractive index-dependent plasmon resonance energy shifts and plasmon-enhanced fluorescence for weak interactions and plasmonic-molecular resonance coupling in the strong coupling limit, leading to hybridized plasmon-molecular states~\cite{Chen:2010}. The interaction behavior is strongly dependent on the electronic structures of the organic molecules. In the first case of weak interactions, the organic molecules do not exhibit light absorption or their electronic absorption energies are far away from the plasmon resonance energies. Their refractive indices are nearly wavelength-independent, and adsorption of these molecules simply increases the refractive index or the dielectric constant of the surface. For the second case -- in the strong coupling regime -- the overall optical response of the hybrid system is the coherent superposition of all interactions, instead of the simple sum of the absorption spectrum from each molecule. All molecules are indirectly coupled via the coherent plasmonic field and behave in the manner of a delocalized macromolecule with notably large coupling strength, thus transforming the response into the collective behavior of the entire molecular ensemble. In this respect, is was shown in the literature that PS\,I can hybridize efficiently with vacuum fields. The photosystem consists of a collection of pigment molecules, such as chlorophylls and carotenoids that are arranged by a protein scaffold in a way that near-field dipole coupling is possible. When interacting with light they no longer act as independent excited molecules, but coupling between them results in collective excitations called excitons, whose wave functions extend over several pigment units. This property results in a very low threshold level to interaction with EM vacuum fields~\cite{Rammler:2022b}. This property can explain therefore the relative large reduction (14 percent) observed for the PS\,I double layer. However, when we consider the full set of results from the various films, we arrive at the conclusion that the high dielectric function and exciton mechanism are probably not the only cause for the reduction since these exciton resonances are restricted to specific narrow absorption bands, while the Casimir effect is influenced by the broadband response of the material. As discussed above however, it is known that the adsorption of molecular films on metal surfaces causes charge transfer between the molecular film and the metal surface resulting in significant surface charge rearrangement that gives rise to a new broad absorption band~\cite{Neuman:2006,Hernando:2006,Carmeli:2003,Hernando:2006a,Carmeli:2003a}. The degree of surface electronic rearrangement increases with the level of organization and dipole moment strength of the molecules. For a single layer, in our experiment, the largest reduction was obtained for PPI. This film has the highest dipole (normal to the surface) in comparison to the C18, PPII, and PS I layers, and was therefore expected to give the strongest effect, as was indeed measured. We therefore link the reduction in the Casimir force to surface charge rearrangement caused by the self-assembly process and the appearance of a broad absorption band in the hybrid gold-monolayer spectra. Further research is required to quantitatively investigate this effect.
\section{Summary and conclusion}
\label{sec:conclusion}
In the present article, we have investigated the reduction of Casimir forces between gold surfaces adsorbed with various types of bio and organic molecular monolayers. The molecules were self-assembled on flat gold surfaces from solution by covalent linking of a thiol group at the end of the molecules. 
This self-assembly resulted in well organized, several nanometer thick monolayers on the surface. We then used a custom-built interferometer-based AFM to measure the force gradient between a gold-coated sphere and the flat gold surfaces coated with molecular monolayers of C18, PP in two conformations, PS\,I single layer, and a PS\,I double layer. For a $4\,$nm monolayer of PP in cis confirmation (PPI), we measured with 5\% more than twice the reduction of the Casimir effect that can obtained with a layer of ITO of the same thickness~\footnote{For the calculation, we used the dielectric functions of Ref.~\cite{deMan:2009zz}.}, while the trans confirmation of PP (PPII) results in no reduction at all. A $9\,$nm monolayer of PS\,I -- being twice as thick as a PP monolayer -- reduces the Casimir interaction by roughly the same amount as ITO. A PS\,I double layer resulted in 14\% reduction.

The present results unambiguously demonstrate the feasibility of the reduction of Casimir forces by surface coatings consisting of self-assembled biomolecular monolayers. In contrast to conventional dielectric materials, these coatings have two major advantages: First, the layer thickness amounts to a mere few nm. Our results with PS\,I indicate that stacking may enhance the effect on the Casimir force significantly, thereby potentially allowing a reduction by several tens of percent with thin layers. The second advantage is the self-assembly property of the investigated layers. These layers may be grown with a well-defined thickness on any kind of substrate in a straight-forward adsorption process from a solution at room temperature. In addition, the small size of the molecules allows them to penetrate nm cavities and trenches. This method therefore appears to be compatible with standard manufacturing procedures in the semiconductor industry, thereby being immediately applicable wherever needed.\\*
We link the observed reduction in the Casimir force gradient to the appearence of a broad absorption band in the gold-monolayer surface, most likely due to charge transfer between the molecular film and the metal. 
As bio and organic molecules are a wide class of materials, further investigations are necessary to fully understand this interaction, and to select the most suitable candidates for an effective practical reduction of Casimir forces. Moreover, as in the present study only one of the interacting surfaces was coated, stronger effects can be expected if both surfaces are equipped with layers of biomolecular monolayers.\\
%
\section{Acknowledgments}
This work was partially funded by the Foundation for Fundamental Research on Matter (FOM) under Projectruimte 10PR2800, and by TU Wien. I. Carmeli acknowledges support under travel grant 040.11.454 from the Netherlands Organization for Scientific Research (NWO). The authors thank H. Abele for administrative support.
\section*{Author contributions}
R. I.P.S: Conceptualization, execution and evaluation of experiments, error calculation, analysis, writing, A.U. spectroscopic measurement analysis, Z.Z.: conceptualization, writing, I.C. conceptualization, preparation of films, analysis, writing.
\section*{Methods}
\subsection*{Experimental Details}
\label{app:details}
\subsubsection*{Dielectric data and models}
\label{app:dielectric}
\begin{table}[!ht]
 \caption{Parameters obtained from least-square fits to the complex dielectric functions of gold and PS\,I.\label{tab:diel_params}}
 \begin{tabular}{cddd|ddd} 
 \hline\hline
     & \multicolumn{3}{c|}{Au} & \multicolumn{3}{c}{PS\,I}\\
     \multicolumn{1}{l}{$\omega_p$ [s$^{-1}$]} & \multicolumn{3}{c|}{$1.2740\times10^{16\phantom{-}\mbox{}}$} & \multicolumn{3}{c}{$1.2008\times10^{16\phantom{-}\mbox{}}$}\\
     \multicolumn{1}{l}{$\tau$ [s]} & \multicolumn{3}{c|}{$1.5549\times10^{-14}$} & \multicolumn{3}{c}{$1.6114\times10^{-14}$}\\
      \hline 
      \multicolumn{1}{c}{$j$} & \multicolumn{1}{c}{$\omega_j$} & \multicolumn{1}{c}{$\xi_j$} & \multicolumn{1}{c|}{$\gamma_j$} & \multicolumn{1}{c}{$\omega_j$} & \multicolumn{1}{c}{$\xi_j$} & \multicolumn{1}{c}{$\gamma_j$}\\
      & \multicolumn{1}{c}{$[10^{15}\,{\rm s}^{-1}]$} & \multicolumn{1}{c}{$[10^{30}]$} & \multicolumn{1}{c|}{[$10^{15}\,{\rm s}^{-1}$]} & \multicolumn{1}{c}{[$10^{15}\,{\rm s}^{-1}$]} & \multicolumn{1}{c}{[$10^{30}$]} & \multicolumn{1}{c}{[$10^{15}\,{\rm s}^{-1}$]}\\
       \hline
     1 & 0.54968& 0.35285 &  0.16020& 0.31401 & 0.18258 & 0.0087736 \\ 
     2 & 0.95685& 1.0957 &   0.82467& 0.32151 & 0.11985 & 0.016133 \\
     3 & 3.1647 &-9.1085 &   2.0516 & 0.55145 & 0.15229 & 0.064765 \\
     4 & 3.6897 &-3.6021 &   0.67997& 0.62680 & 0.11765 & 0.039870 \\
     5 & 4.0210 & 7.6035 &   1.0930 & 0.96289 & 1.4576 &  0.73387 \\
     6 & 4.6597 & 26.598 &   1.8806 & 2.9987 & -3.7963 &  1.6237 \\
     7 & 5.7443 & 10.835 &   1.2756 & 3.6564 & -2.8441 &  0.74669 \\
     8 & 6.6361 & 47.435 &   2.5470 & 4.0978 &  7.9538 &  1.0414 \\
     9 & 8.4053 & 5.9287 &   0.35794& 4.7613 & 23.753 &   1.7940 \\
     10& 12.967 & 151.04 &   6.5412 & 5.6641 &  8.4203 &  1.0933 \\
     11& 20.089 & 634.95 &  57.231  & 6.4625 & 28.337 &   1.9244 \\
     12& 31.996 & 211.73 &   6.4693 & 7.6733 &  3.5074 &  1.9145 \\
     13& 84.669 & 1665.2 & 119.99   & 12.689 & 206.24 &   8.5861 \\
     14& 44.991 & 482.32 &  24.901  & 15.584 & 162.40 &  12.897 \\
     15& 22.836 & 35.404 &  -7.2073 & 18.880 &  57.704 & 15.193 \\
     16& 20.275 & 205.32 &  10.094  & 31.227 & 276.97 & 8.2679 \\
     17& 69.799 &-2124.9 &-420.94   & 41.095 & 383.15 & 30.590 \\
     18& 225.58 &-2578.3 & 252.88   & 51.658 & 695.11 & 40.903 \\
     19& 349.80 &-3890.7 &-441.04   & 97.676 & 385.50 & 45.814 \\
     20& 699.82 & 671.57 & 191.05   & 19.532 & -83.292 & -38.195 \\
     21& 1180.6 & 2553. & 920.96    & 100.38 & -2577.5 & -264.41 \\
     22& 3956.7 & 830.09 & 740.25   & 265.59 & -2913.8 & 322.23 \\
     23& 5280.2 & 2599.5 &3713.0    & 351.98 & -3110.5 & -305.28 \\
     24&  {-}   &  {-}   &   {-}    &  669.17 & 2148.0 & 412.18 \\
     25&  {-}   &  {-}   &   {-}    &  1232.8 & 2654.0 & 951.40 \\
     26&  {-}   &  {-}   &   {-}    &  4109.3 & 1676.6 & 1377.5 \\
     27&  {-}   &  {-}   &   {-}    &  5932.3 & 1856.4 & 4193.8 \\
     \hline\hline
    \end{tabular}
 \end{table}      
The spectroscopic data on $\psi$ and $\Delta$~\footnote{The angles $\psi$ and $\Delta$ are related to the reflection coefficients $r_p$ and $r_s$ in and out of the plane of incidence, respectively, by $\rho=r_p/r_s=\tan\psi\exp{\ri\Delta}$. $\psi$ and $\Delta$ are obtained by modeling data on $\rho(\phi)$ at different angles of incidence $\phi$~\cite{Woollam:1999}.} as well as the complex dielectric functions~\footnote{The complex dielectric function $\vare=\vare_1+\ri\vare_2$ is related to the measured ratio $\rho(\phi)$ by $\vare=\sin^2\phi\left[1+\tan^2\phi(1-\rho)^2/(1+\rho)^2\right]$~\cite{Woollam:1999}.} were obtained using commercial RS2 and IR-VASE ellipsometers and the VWASE software by Woollam Inc. In the conversion, a single effective layer was assumed instead of a thin layer atop a bulk material. This choice is rectified by the hybrid electron states creating an effective material response. We performed a synchronous least-squares fit of the double-logarithmically scaled dielectric functions $\log \vare(\log\omega)$ for Au and PS\,I to the model given in \eqnref{eq:Drude}. In doing so, we appended our own datasets at $\omega>7.53\times 10^{15}\,s^{-1}$ by literature values~\cite{Palik:1998a} for gold. The parameters resulting from the fits are given in \tabref{tab:diel_params}. Note that the obtained plasma frequency for gold differs from the frequently used value $1.3673\times 10^{16}\,{\rm s}^{-1}$ given for a \emph{thin} film in the literature~\cite{Palik:1998a}. This fact highlights the importance of spectroscopic sample characterization for precise comparisons of experiment and theory.
\subsubsection*{Data Handling and Error Propagation}
\label{app:data-handling}
We analyze our gradient measurement data in several steps. First, we repeat the determination of the unknown shift $a_0$ that offsets the recorded calibrated piezo positions $a_{pz}$ from the actual separation $a$ for each sweep by means of a fitting procedure documented in the literature~\cite{deMan:2009}. Thermal (and other) drift $\Delta a_0(i)=a_0(i)-a_0(i-1)$ between subsequent sweeps (indexed here by $i$) increase the uncertainty in $a_0$. In order to limit the influence of this systematic uncertainty on our data, we disregard any sweeps with $|\Delta a_0|>5\,$nm in the evaluation. We use a spline interpolation between all remaining subsequent $a_0$, assumed to be exact at the center points of each sweep (point number 16). In this way, we compute the most probable actual $a_0$ for each point in a sweep using the recorded times of the points as ordinate. This interpolation removes the effect of constant drifts entirely but may give erroneous results in the case of rarely occurring faster variations, for which we further discard any runs having $|\Delta a_0(i)-\Delta a_0(i-1)|>3\,$nm. The error in the resulting $a$ for each point consists of the propagated statistical errors of the applied voltages $V_{AC}$ and $V_{ex}$, the signal amplitude $S_{2\omega1}$, and the uncertainty in $a_0$ and $a_{pz}$. We apply a similar interpolation technique to compensate for the minor changes in $\omega_0$, which is determined before starting each sweep. Recorded variations of the resonance frequency are most probably due to indirect effects of thermal fluctuations in the air-conditioned laboratory (outside the anechoic box $19\pm1\,^\circ$C, resulting in variations of $\lesssim10$ \% in humidity), and have never exceeded $80\,$mHz over a typical duration of 48 hours per run. After correction of $\omega_0$ for each point, the total force gradient is computed according to \eqnref{eq:force_gradient}, and reduced by the known gradient due to the applied electrostatic excitations $V_{AC}$ and $V_{ex}$ to obtain the Casimir force gradient. Statistical errors from all voltage measurements, the PLL frequency determination, and the calibrations of $m$, and $\omega_0$ (see section on calibration below) are propagated to the final result. 
In order to give a more complete picture, we present our (raw) data at this evaluation stage in \figref{fig:raw_data}.
 \begin{figure}[!ht]
  \centering
  \includegraphics[scale=0.92]{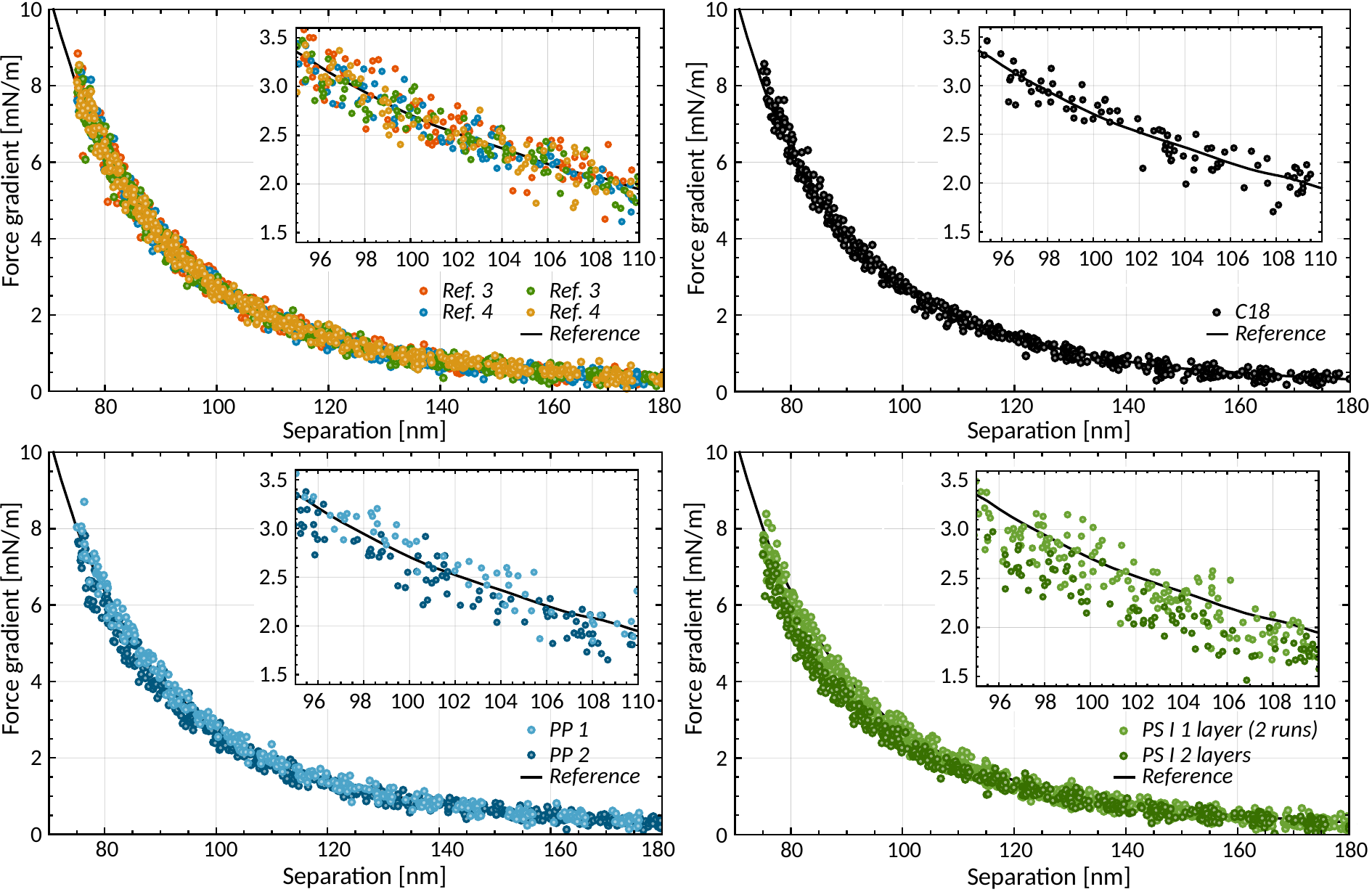}
  \caption{Measured force gradients for all five substrates obtained from the same data as the results in \figref{fig:results}, but without interpolation of $d_0$, $\omega_0$, discrimination of sweeps with large $\Delta d_0$, and averaging.\label{fig:raw_data}}
 \end{figure}
Eventually, we apply a running mean with width equal to the number of sweeps taken into account for the present evaluation, weighted by the inverse square of the errors in $a$ (in horizontal direction) and $\partial_a F_C$ (vertical direction)~\footnote{Note that the same averaging procedure has been used in the literature~\cite{Intravaia:2013}.}. The resulting error of the mean is shown together with the averaged data in the main text in \figref{fig:results}.
\subsubsection*{Calibration and Error estimation}
\label{app:cal}
In order to determine the resonance frequency and phase of the cantilever, we increase the separation to $a=a_{cal}\approx\um{2.5}$, perform a sweep of the excitation frequency $\n_{2}$ around $\omega_0/4\pi$~\footnote{Note that the frequency of the electrostatic force exciting the cantilever is twice the one of the applied electrostatic signal.}, and record the phase of the optically measured mechanical response of the cantilever with respect to the second harmonic of the excitation. The value of $\omega_0$ is then extracted by fitting the argument $\mathop{Arg}\left[F_{ES}(a,V_{ex},V_{AC})T_{Fx}\right]+\phi_{off}$, with the cantilever response function
\begin{align}
 F_{ES}(a,V_{ex},V_{AC})T_{Fx}=\frac{4\pi R\vare_0\left(V_{ex}^2+V_{AC}^2\right)}{a^2 \left(m\omega_0^2+\left[2\pi\vare_0R^2\left(V_{ex}^2+V_{AC}^2\right)\right]/{a^2}+\ri \gamma\omega- \omega^2\right)}\,.\label{eq:cal_tf}
\end{align}
to the data. After completion of the frequency sweep, we deactivate the $V_0$ feedback circuit, set $\n_{2}=\omega/4\pi$, and record the frequency shift for a sweep of $V_{DC}$. From the resulting parabola, we can determine $m$, $V_0$, and any error $\omega_{off}$ of $\omega_0$ by means of a least-squares fit of \eqnref{eq:cal_tf} to the data. While $m=(1.871\pm0.036)\times10^{-8}\,$kg stayed constant within the standard deviation over six months of measurements, the resonance frequency decreased gradually from $609.07\,$Hz to $609.01\,$Hz during this time.
The effective sphere radius on the tip (region of closest approach) was determined after the end of the measurements by fitting a spherical surface to $10\times 10\si{\micro\square\metre}$ AFM scans taken with two different calibrated instruments. Details of this procedure can be found in Ref.~\cite{Sedmik:2013}. We obtained a value $77.9\pm0.8\,\si{\micro\metre}$ that compares well with the nominal NIST-certified radius of $79.5\pm2.2\,\si{\micro\metre}$.

\begin{table}[!hb]
 \caption{Indicative contributions of various error sources to the obtained force gradients normalized by the sphere radius at a separation of $100\,$nm for the gold reference sample, excluding roughness and patch potentials. These values need to be compared to the measured Casimir force gradient $\partial_a F_C(\SI{100}{\nano\metre})=\SI{2.754}{\milli\newton/\metre}$ for the Au references.\label{tab:errors}}
 \begin{tabular}{l r @{.}l}
 \hline\hline
  Type of uncertainty &\multicolumn{2}{c}{contribution [\si{\micro\newton/\metre}]}\\
  \hline
  sphere radius & 9&4\\
  effective mass & 78&7\\
  applied voltages & 0&05\\
  frequency measurement &78&4\\
  resonance frequency & 31&8\\
  distance determination &16&6\\
  \hline
  total statistical &116&1\\
  total systematic &15&3\\
  \hline
 \end{tabular}
\end{table}
As can be seen in \tabref{tab:errors}, the major statistical contributions come from the mass calibration and the frequency measurement. These errors could be reduced by increasing the integration time per point, which, however, would also lead to increased drift (and hence systematic errors) between subsequent sweeps.

An important systematic effect plaguing most Casimir force measurements is the so-called `patch effect' caused by locally varying Fermi potentials due to crystal grains and impurities. In order to estimate the influence of this effect, we have performed Kelvin probe force microscopy on each of the investigated substrates using a modified Bruker Nanoscope III device~\cite{Polak:2014} in AM mode (see \figref{fig:topopot} in the main text). 
We followed the method of Ref.~\cite{Behunin:2012zr} to compute the electrostatic interaction energy $E_{sp}$ according to their Eq. (13) without self-interactions and under consideration of the globally minimizing potential $V_0$. Although the potential distributions of both surfaces have been measured, it is impossible to compute the patch contribution exactly, as the resulting force(gradient) depends on the relative position of sphere and plate. In order to quantify this effect, we created numerically a repeated tiling of the measured potential distribution on the substrates and selected $N_{MC}>100$ random positions on the resulting surface for the the point of closest approach. The corresponding patch energy for the $N_{MC}$ configurations were then averaged to result in the most probable actual patch contributions. The statistical spread between the different solutions, however, vastly overestimates the actual uncertainty for the situation in the experiment. This is due to the applied selection procedure for the position for data taking. Any local corrugations or potential maxima would lead to snap-to-contact at nominal surface separations $>50\,$nm. As for the force gradient measurements we specifically selected positions at which we can approach separations much shorter than that, we have experimentally filtered out locations with strong disturbances in both roughness and surface potential. While it may hardly be possible to re-create the very same conditions for our numerical estimates, we exclude any position at which the minimum distance between the rough opposing surfaces reduces below 30 nm, and further reject runs with excessive numerical integration error, manifesting itself in steps or outliers in the computed energy as a function of surface separation. We note at this point that the numerical evaluation of the patch contribution according to Ref.~\cite{Behunin:2012zr} is challenging and time consuming. Eventually, we include 31 runs for gold and 22 for PS\,I in the evaluation, and numerically compute the force gradient $\partial_a F_{\text{patch}}(a)$ from the previously obtained patch energy. For these results, we compute the 68.3\% ($1\,\sigma$) probability ranges based on the distribution from our Monte Carlo computation for each separation, for gold and PS\,I.

We apply a similar procedure for the roughness contribution where, due to the lower time effort for computations in comparison to the patch potential contributions, we continued to generate new shifts until $N_{MC}=100$ acceptable runs were found. Note that for the sphere side, we used a larger AFM dataset than the one shown in \figref{fig:topopot} with side length \SI{25}{\micro\metre} that permitted to remove from it a spherical surface fit with radius $R=\SI{77.9}{\micro\metre}$ and adjust the point of closest approach to the most probable location on the resulting flat surface, which showed minor local flattening of the profile, most probably due to repeated gentle snap-over. For the roughness computations, we added again the spherical surface centered at the point of closest approach, and took into account the fully patterned spherically deformed surface.
We used the proximity force approximation~\cite{Derjaguin:1956,Bordag:2014} to compute for 100 random sets of lateral shifts $(x_o,y_o)$ of the flat plate the relative roughness correction $\eta_{\text{rough},M}(a,x_o,y_o)=(1/N)\iint\!{\rm d}r\, {\rm d}\phi\,\partial_a F_C[a+R-\sqrt{R^2-r^2}-r_s(r,\phi)-r_{p,M}(r,\phi)]$, where $N$ is the Casimir force gradient at nominal separation $a$ between perfectly smooth surfaces, and $r_s$ and $r_p$ are the measured roughness profiles on the sphere and flat substrates, respectively. Again, we determined the 68.3\% probability ranges at each separation based on the statistical spread between the various shifts at each $a$, and the uncertainty in $R$.

Next, we computed the correction factors $\eta_{\text{rough},M}(a) = \partial_a F_{C,\text{rough},M}(a)/\partial_a F_{C,\text{flat},M}(a)$ between rough and flat surfaces, and correspondingly $\eta_{\text{patch},M}(a) = \partial_a F_{\text{patch},M}(a)/\partial_a F_{C,\text{flat},M}(a)$ for each material $M$, considering the respective uncertainties. These data allow us to compute the relative corrections $\eta_{\text{rough,rel}}(a)=\eta_{\text{rough},\text{PS\,I}}(a)/\eta_{\text{rough},\text{Au}}(a)$ and $\eta_{\text{patch,rel}}(a)=\eta_{\text{patch},\text{PS\,I}}(a)/\eta_{\text{patch},\text{Au}}(a)$, which we show separately and combined in  \figref{fig:corrections}. The blue line only considers only the change due to the measured dielectric properties [$\partial_a F_{C,\text{rel}}(a)=\partial_a F_{C,\text{PSI}}(a)/\partial_a F_{C,\text{Au}}(a)-1$], while the orange and green curves show  $\partial_a F_{C,\text{rel}}(a) (\eta_{\text{rough,rel}}(a)-1)$ and $\partial_a F_{C,\text{rel}}(a) (\eta_{\text{rough,rel}}(a) \eta_{\text{patch,rel}}(a)-1)$, respectively. As expected, both corrections increase the force gradient on PS\,I with respect to gold. The uncertainties from the Monte Carlo computations of the patch contribution are relatively large, resulting in the average force gradient reduction of $\Delta_\text{PS\,I}[\partial_a F_\text{tot}](a)\equiv \partial_a F_{C,\text{rel}}(a)(\eta_{\text{rough,rel}}(a)\eta_{\text{patch,rel}}(a)-1)=-2.7${\raisebox{0.5ex}{\tiny$\substack{+1.2 \\ -1.7}$}\% for the range 80--120\,nm given in the main text. The green curve is also included in \figref{fig:results} for reference. 
\begin{figure}[!ht]
\centering
\includegraphics{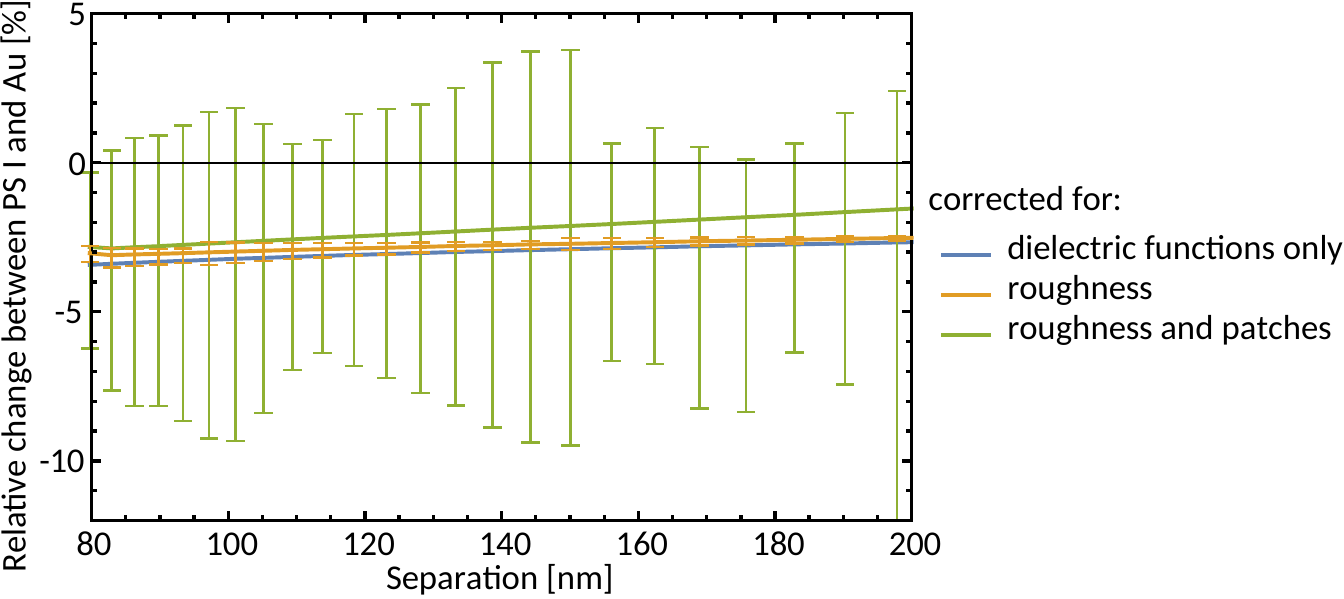}
\caption{Monte Carlo results for the relative corrections due to patch potentials and roughness for different lateral shifts of the interacting surfaces. The error bars indicate the 68.3\% probability ranges. \label{fig:corrections}}
\end{figure}

Despite eliminating runs with too small minimum surface separations, we think that the theoretical procedure described here still over-estimates the real error (at least for the patch potential contribution). This conclusion is supported by the fact that force gradient measurements on three different PS\,I substrates agreed within the experimental standard deviations of 1--2\% given in \tabref{tab:reduct} in the main text. The latter fact gives confidence that patches, which may differ between locations and samples, do not have a strong influence on our results. Moreover, as noted in the main text, stronger electrostatic effects would lead to an increase in the force rather than a decrease, as is observed here.

\subsection*{Frequency shift measurements in air}
Measurements of (Casimir) force gradients are usually performed in vacuum~\cite{Decca:2003zz,Decca:2007jq,Chang:2012fh,Kawai:2016}, as the mechanical Q-factor of the cantilever can be maximized and only in-phase force gradients need to be considered~\cite{Kleiman:1985}. On the other hand, in liquids the strong influence of viscous and kinetic forces can be used to extract hydrodynamic parameters from the measured frequency shift and phase~(examples: \cite{Xu:2006,Ma:2000}, slightly outdated review:~\cite{Butt:2005}). The measurements presented here are performed in an intermediate regime in which kinetic forces do not play a role but viscous squeeze film damping could influence the results. 
\begin{figure}[!hb]
 \centering
 \includegraphics[scale=0.68]{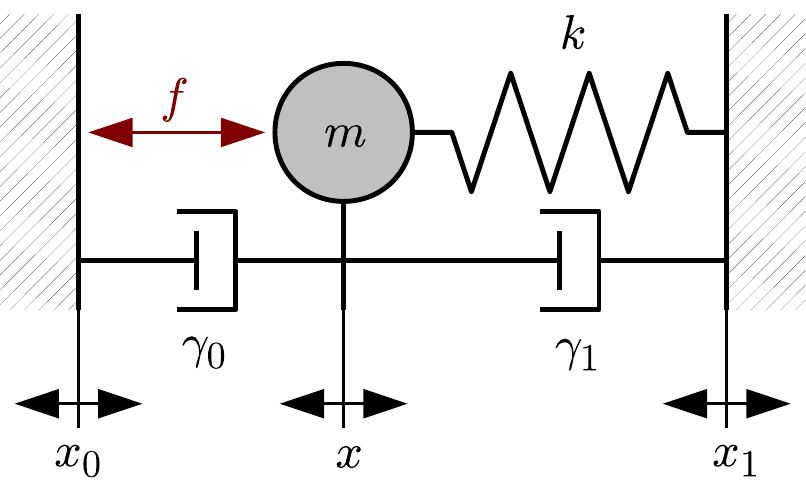}
 \caption{Specification of the geometric parts used in the lumped parameter model \eqnref{eq:eom_moving_base}. We have damping coefficients $\gamma_i$, positions $x_i$, the force $f$ acting between the left plate and the sphere on the cantilever with mass $m$, suspended with spring constant $k$\label{fig:lumped_param_model}.}
\end{figure}
We model the cantilever using lumped parameters as shown in \figref{fig:lumped_param_model}, with an oscillator of mass $m$, spring constant $k$, and viscous dampings $\g_0$ towards the substrate (left) and $\gamma_1$ towards the base (right). This system obeys the equation of motion
\begin{align}
 m\partial_t^2x(t)+\gamma_0\partial_t \left[x(t)-x_0(t)\right]+\gamma_1\partial_t \left[x(t)-x_1(t)\right]+k\left[x(t)-x_1(t)\right]=f(t)\,.\label{eq:eom_moving_base}
\end{align}
While $\g_1$ is likely to be fixed for even moderate deformations of the cantilever, $\g_0$ may depend on the separation $a$ between (the probe on) the cantilever, and the opposing substrate (plate). In that case, and as long as the movement of the (gaseous) medium around the cantilever results in only viscous forces~\footnote{Elastic effects can be excluded if the so-called \emph{squeeze number}~\cite{Maali:2008} and the Reynolds number both are $\ll1$ for the experimental conditions encountered here.}, $\g_0$ in \eqnref{eq:eom_moving_base} takes the form $C(a-x)^{-n}$ with $C$ being a geometric constant, and $n$ being the exponent of the force. For a spherical probe, $\g_0\approxprop a^{-1}$~\footnote{Strictly speaking, for ranges $0.01\lesssim Kn<1$ of the Knudsen number $Kn=\lambda_f/a$ and a mean free path $\lambda_f\approx 70\,$nm for air at ambient pressure, slip effects need to be taken into account, resulting in $n<1$. For the qualitative argument discussed here, however, this detail may be neglected. A more in-depth discussion is given in~\cite{Sedmik:2013}}. We expand $\gamma_0$ for small movements $\Delta a=x-x_0$ around the equilibrium point $a$,
\begin{align}
 \gamma_0(a-\Delta a)\underset{|\Delta a|\ll a}{=}\gamma_0(a)-[\partial_a\gamma_0(a)]\Delta a+\mathcal{O}\left(\Delta a^2\right)\,,
\end{align}
and similarly for the force $f(a-\Delta a){=}f(a)-[\partial_af(a)]\Delta a+\mathcal{O}\left(\Delta a^2\right)$. Then, we insert these expressions for $\gamma_0(a-\Delta a)$ and $f(a-\Delta a)$ into \eqnref{eq:eom_moving_base}, and eliminate $x$ from the equation by defining $x=y+x_1$. This last step is motivated by the fact that in experiments it is often more convenient to measure the relative movement $y$ between cantilever and base than the absolute movement $x$. We thus arrive at the modified equation of motion (where we omit the explicit time dependences)
\begin{align}
 m\partial_t^2(y+x_1)+\gamma_0\partial_t \left(y+x_1-x_0\right)-(\partial_a\gamma_0)\left(y+x_1-x_0\right)\partial_t\left(y+x_1-x_0\right)+\gamma_1\partial_t y +\left(k-\partial_a f\right)y=f\,.\label{eq:eom_moving_base_complete}
\end{align}
Assuming as usual purely sinusoidal excitations $x_0(t)$, $x_1(t)$, and $f(t)$, \eqnref{eq:eom_moving_base_complete} can conveniently be solved in Fourier space, leading to
\begin{align}
Y(\omega)= \frac{F-(\partial_a F)(X_0-X_1)-\ri\omega\left(\gamma_0(X_1-X_0)+3(\partial_a\gamma_0)X_0X_1+\ri\omega m X_1\right)}{k-\partial_aF+\ri\omega\left(\gamma_0+\gamma_1+2(\partial_a\gamma_0)(X_0-X_1)\right)-m\omega^2}\,,\label{eq:solution_y_complete}
\end{align}
where the capital letters $B={F,X_0,X_1,Y}$ denote the Fourier-transformed versions $B(\omega)=\left(2\pi\right)^{-1/2}\int_{-\infty}^{\infty}\!{\rm d}t\,b(t)\exp\left(-\ri\omega t\right)$ of the respective variables $b$ in non-capital letters in \eqnref{eq:eom_moving_base_complete}.
As \eqnref{eq:solution_y_complete} does not permit the simple definition of transfer functions, we proceed by expanding $Y(\omega)$ for small amplitudes~\footnote{We assume $F(\omega,a)=F_0(a)+\Delta F(\omega)$ with $\Delta F(\omega)\ll1$, $X_0(\omega)\ll1$, and $X_1(\omega)\ll1$.} for each of the three excitations $B$ (except for $Y$) separately up to first order. The linear coefficients $\tilde{Y}_B\equiv\left.\partial Y/\partial B\right|_{B\to 0}$ of the first order then give an approximation of the transfer functions $T_{YB}(\omega)$ that describe the small signal response of the relative movement $Y$ to excitation $B$. 

Having in mind the application of a PLL, we strive to determine the resonance frequency $\omega_{0,B}$ from the phase jump by $\pi$ occurring at the resonance. Therefore, we use the condition $\vp(\omega_R)=\vp(0)-\pi/2$. We note, however, that for excitation via $X_1$ and $X_0$, $\gamma_0$ and $\partial_a \gamma_0$ cause additional phase offsets that shift $\vp$ with non-trivial dependence on $\omega$ invalidating the condition for $\vp(\omega_{0,B})$. As these additional offsets vanish for $\gamma_0,\,\partial_a\gamma_0\to0$, we nonetheless resort to solving ${\rm Re}\tilde{Y}_B(\omega_{0,B})=0$ for $\omega_R$ for all excitations. We then determine the corresponding phase according to $\vp_{0,B}[\tilde{Y}_B(\omega_{0,B}]=\arctan\left[{\rm Im}\,Y(\omega_{0,B})/{\rm Re}\,Y(\omega_{0,B})\right]$. In this way, we obtain~\footnote{Note the notation that $\partial_a f$ denotes the derivative of the absolute force acting over $a$, while $F$ is the amplitude of a sinusoidal excitation in frequency space. Note further that notationwise the derivatives $\partial_a$ act exclusively on the $f$ or $\gamma_0$ following immediately after.}.
\begin{align}
 \tilde{Y}_F: &\quad\omega_{0,F}=\sqrt{\frac{k-\partial_a f}{m}}\, &&\vp_{0,F}=-\frac{\pi}{2}\,,\label{eq:omega_0_res_F}\\
 \tilde{Y}_{X0}: &\quad\omega_{0,X0}=\sqrt{\frac{k-\partial_a f}{m}}\, &&\vp_{0,X0}=\arctan\frac{-\sqrt{m(k-\partial_a f})\left[\partial_a f(\gamma_0 + \gamma_1) + 2\partial_a\gamma_0 (F + (k-\partial_a f) X0)\right]}{(\partial_a f - k) \left[\gamma_0 (\gamma_0 + \gamma_1) -2 \partial_a\gamma_0 (\gamma_0 + \gamma_1) X1 + 4 (\partial_a\gamma_0)^2 X1^2\right]}\,,\label{eq:omega_0_res_X0}\\
 \tilde{Y}_{X1}:&\quad\omega_{0,X1}=\sqrt{\frac{k-\partial_a f}{m}}\, &&\vp_{0,X1}=\arctan\frac{-\sqrt{m(k-\partial_a f})\left[k(\gamma_0 + \gamma_1) + 2\partial_a\gamma_0 (F + (k-\partial_a f) X0)\right]}{(\partial_a f - k) \left[\gamma_0 (\gamma_0 + \gamma_1) +2 \partial_a\gamma_0 (\gamma_0 + \gamma_1) X0 + 4 (\partial_a\gamma_0)^2 X0^2\right]}\,.\label{eq:omega_0_res_X1}
\end{align}
Note that we have checked numerically that there exist no $\omega_{0,X0}$ or $\omega_{0,X1}$, for which a phase could be obtained that is constant under variation of $a$, i.e. under changes of $\gamma_0$, $\partial_a\gamma_0$, and $F$. If we set all other sources $C\neq B$ to zero, \eqnref{eq:omega_0_res_X0} and \eqnref{eq:omega_0_res_X1} simplify to
\begin{align}
\lim_{F,X1\to 0}{\vp_{0,X0}}={\atan}\frac{\partial_a fm}{\gamma_0\sqrt{m(k-\partial_a f)}}\,,\quad\text{and }\lim_{F,X0\to 0}{\vp_{0,X1}}={\atan}\frac{km}{\gamma_0\sqrt{m(k-\partial_a f)}}\,,
\end{align}
which still have an explicit dependence on distance. Only $\omega_{0,F}$ and $\vp_{0,F}$ stay unaffected by changes of $a$. As for measurements in gas $\gamma_0$ will always be non-zero, force excitation ($\tilde{Y}_F$) is the only way to perform accurate measurements of force gradients using a PLL. For measurements of the absolute movement $x(t)$ the same qualitative conclusion holds.
%

\end{document}